\documentclass[reprint,
twocolumn,
 amsmath,amssymb,
 aps
]{revtex4-1}

\usepackage{graphicx}
\usepackage{dcolumn}
\usepackage{bm}
\usepackage{float}
\usepackage{xcolor}
\begin{document}

\title{Asymmetry of forward/backward transition times as a non-equilibrium measure of complexity of microscopic mechanisms}

\author{Jaeoh Shin$^{1,2}$}
\author{Anatoly B. Kolomeisky$^{1,2,3,4}$}
\affiliation{$^1$Department of Chemistry, Rice University, Houston, Texas, 77005, USA}
\affiliation{$^2$Center for Theoretical Biological Physics, Rice University, Houston, Texas, 77005, USA}
\affiliation{$^3$Department of Chemical and Biomolecular Engineering, Rice University, Houston, Texas, 77005, USA}
\affiliation{$^4$Department of Physics and Astronomy, Rice University, Houston, Texas, 77005, USA}

\begin{abstract}
In one-dimensional random walks, the waiting time for each direction transitions is the same, even in the presence of bias, as a consequence of the microscopic-reversibility. We study the symmetry breaking of forward/ backward transition times in a random walk on a lattice of two lanes. We find that either transition times can be faster depending on the lattice's net current, and the symmetry is recovered only at equilibrium. Our analysis suggests that the forward/ backward transition times' asymmetry can be used as a measure of deviation from the equilibrium of the system.
\end{abstract}

\maketitle


One of the main experimental tools in studying the microscopic mechanisms of complex phenomena is measuring transition times for individual particles \cite{thorneywork2020direct}. These observations are frequently analyzed using a random walk approach, which is one the simplest theoretical models that accounts for stochasticity at the molecular level \cite{berg1993random}. In the presence of a bias, a random walker moves preferentially along  the bias; however, the jumps' waiting time is the same in both directions as a consequence of  microscopic reversibility \cite{qian2006open}. In has been argued recently  that the symmetry of the downhill (along with the bias) and uphill (against the bias) transition times can be broken in multi-particle systems with interactions \cite{ryabov2019counterintuitive, shin2020biased}. Here we show that such symmetry might not be observed at non-equilibrium conditions even for a single particle due to a complexity of the underlying molecular mechanisms.

\begin{figure}
    \centering
    \includegraphics[width=0.7\columnwidth]{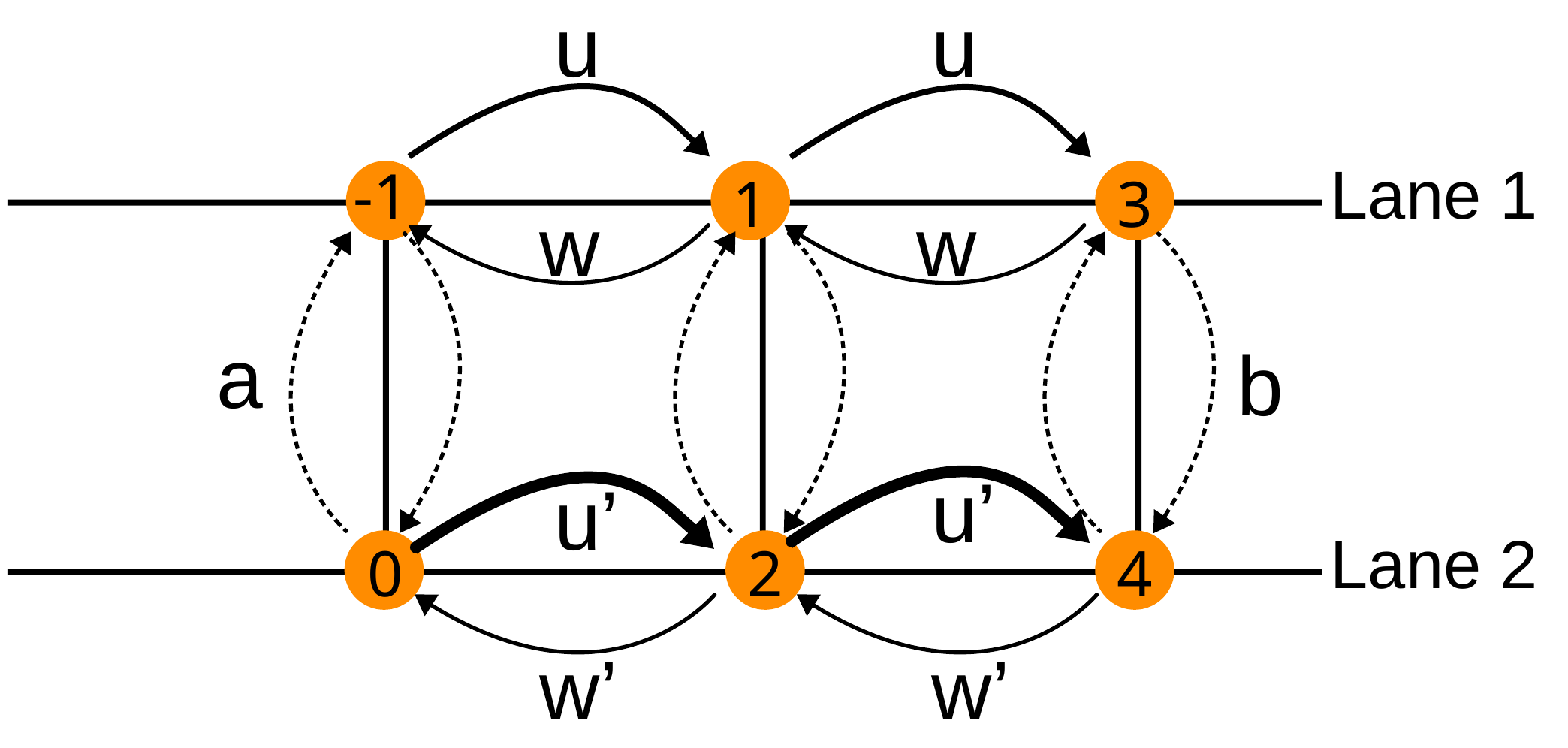}
    \caption{A model for a single particle hopping on a two-lane lattice. The transition rates are $u$ and $w$ to move to the right and to the left, respectively, in the lane 1 lane; and the corresponding rates in the lane 2 are $u'$ and $w'$. The particle can also jump between the lanes with the rates $a$ and $b$.}
    \label{Fig-1}
\end{figure}

Consider a particle moving on a two-lane lattice as illustrated in Fig.\ref{Fig-1}. In lane 1, the hopping rates are $u$ and $w$ to the right and left, respectively. In lane 2,  $u'$ and $w'$ are the corresponding right and left rates. The particle can also transition between two lanes with the rate $a$ (from lane 2 to lane 1) and $b$ (from lane 1 to lane 2). The two-lane system in Fig. \ref{Fig-1} can be viewed as a model for a motor protein moving along the linear track when the protein can be found in one of two molecular conformations \cite{kolomeisky2015motor}. We define the forward motion as the motion to the right along any lane, and, correspondingly, the backward motion is any motion to the left.

We start with the transition to the right through lane 1. Since the transition time is defined only between two neighboring sites, we consider  six sites on the lattice  marked  $-1,0,1,2,3,4$ in Fig. \ref{Fig-1}. It is assumed that the particle is placed either at the site 1 or at the site 2 at time $t=0$, and the system is transitionally invariant. Now we define first-passage probability-density distribution functions $F_{i,3}(t)$ ($i=1,2,3$) to reach the state $3$  starting at the site $i$ at time $t$ before making any forward/backward transition via other paths. These functions can be found from the backward master equations \cite{redner2001guide,kolomeisky2015motor},
\begin{equation}
\frac{dF_{1,3}(t)}{dt}=-(u+w+b)F_{1,3}(t)+bF_{2,3}(t)+uF_{3,3}(t),
\end{equation}
\begin{equation}
\frac{dF_{2,3}(t)}{dt}=-(u'+w'+a)F_{2,3}(t)+aF_{1,3}(t),
\end{equation}
and $F_{3,3}(t)=\delta(t)$.
Applying the Laplace transformation, $\widetilde{F}_{i,3}(s)\equiv \int_{0}^{\infty} F_{i,3}(t)\exp(-st)dt$, we derive
\begin{equation}
(s+u+w+b)\widetilde{F}_{1,3}(s)=b\widetilde{F}_{2,3}(s)+u,
\end{equation}
\begin{equation}
(s+u'+w'+a)\widetilde{F}_{2,3}(s)=a\widetilde{F}_{1,3}(s).
\end{equation}
These  equations can be easily solved, allowing us to obtain  the  splitting probabilities to reach the state 3 before reaching other states ($-1,0,4$) starting from the state 1, $\Pi_{1,3}$, or starting from the state $2$, $\Pi_{2,3}$:
\begin{equation}
\Pi_{1,3}=\frac{u (a + u' + w')}{a (u + w) + (b + u + w) (u' + w')},
\end{equation}
\begin{equation}
\Pi_{2,3}=\frac{au}{a (u + w) + (b + u + w) (u' + w')}.    
\end{equation}
The (conditional) mean-first passage times $T_{1,3}, T_{2,3}$ for such events can be also found from the first-passage distributions \cite{redner2001guide,kolomeisky2015motor}, producing
\begin{equation}
T_{1,3}=\frac{a^2 + (u' + w')^2 +  a (b + 2 (u' + w'))}{(a + u'+w') [a (u + w) + (b + u + w) (u' + w')]},
\end{equation}
\begin{equation}
T_{2,3}=\frac{a + b + u + u' + w + w'}{a (u + w) + (b + u + w) (u' + w')}.   
\end{equation}
Similarly, we determine the forward transition times via the lane 2 by evaluating the first-passage  distribution functions $F_{1,4}(t)$ and $F_{2,4}(t)$ to reach the state 4, yielding
\begin{equation}
\Pi_{1,4}=\frac{b u'}{a (u + w) + (b + u + w) (u' + w')},
\end{equation}
\begin{equation}
\Pi_{2,4}=\frac{u' (b + u + w)}{a (u + w) + (b + u + w) (u' + w')};
\end{equation}
\begin{equation}
T_{1,4}=\frac{a + b + u + u' + w + w'}{a (u + w) + (b + u + w) (u' + w')}, 
\end{equation}
\begin{equation}
T_{2,4}=\frac{a b + (b + u + w)^2}{(b + u + w) [a (u + w) + (b + u + w) (u' + w')]}.
\end{equation}

Finally, the mean forward transition time can be calculated by averaging over all the possible forward paths and initial states,
\begin{equation}
T_{+}(u,w,u',w')= \frac {\sum_{i=1,2}\sum_{j=3,4} p_i \Pi_{i,j}T_{i,j}}{\sum_{i=1,2}\sum_{j=3,4} p_i \Pi_{i,j}}    
\end{equation}
where $p_i$ ($i=1,2$) is the probability for the particle to start at the lane 1 or lane 2, respectively. Measuring the transition events from  long-time trajectories \cite{shin2020biased} gives $p_1=\frac{a}{a+b}$ and $p_2=1-p_1$. The backward transition times can be obtained from symmetry considerations by exchanging $u$ and $u'$, and $w$ and $w'$, $T_{-}(u,w,u',w')=T_{+}(w,u,w',u')$.

\begin{figure}
    \centering
    \includegraphics[width=0.7\columnwidth]{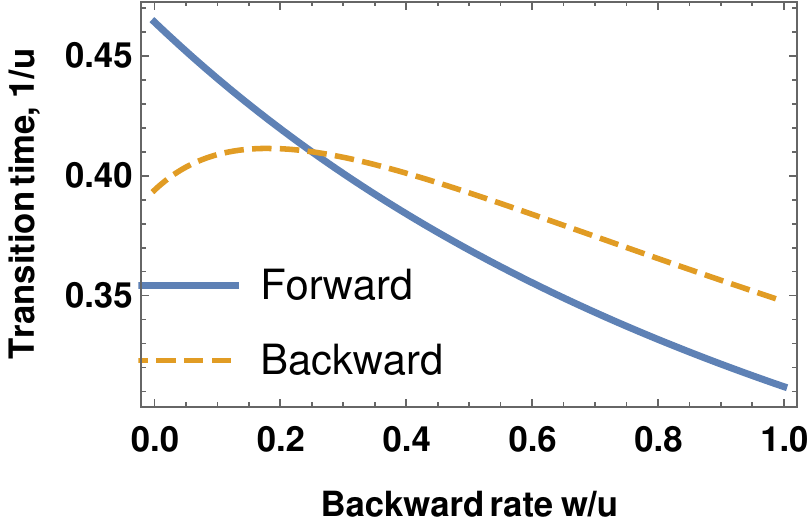}
    \caption{Forward and backward transition times. The following parameters are used in calculations: $u=1$, $u'=4$, $w'=1$, $a=b=1$. Only the rate $w$ is varied.}
    \label{Fig-2}
\end{figure}

We found that the transition times of forward and backward are different in general. As an illustration, we show the forward and backward transition times in Fig. \ref{Fig-2} for a specific set of parameters. Surprisingly, either forward or backward transition time can be faster even though at the given conditions the particle prefers to move to the right. 

To explain these observations, consider a cyclic network of states $(1,3,4,2)$. The thermodynamic affinity for this loop, which can be viewed as a driving force for the system to be out of equilibrium, can be evaluated as \cite{qian2006open,hill2005free}, 
\begin{equation}
A=k_{\text{B}}T \ln \left( \frac{u b w'a}{wbu'a}\right)=k_{\text{B}}T \ln \left(\frac{uw'}{u'w}\right).
\end{equation}
Only at equilibrium we have $A=0$, and $A \neq 0$ corresponds to a net current in the system. With the choice of our parameters in Fig. 2, we have $A>0$ ($<0$) for $w<1/4$ ($>1/4$). Fig. 2 shows that for the clockwise current ($A>0$ and $w<1/4$), the forward transition time is longer than the backward transition time. The particle preferentially moves forward through the lane 1, where the transition is slower than through the lane 2. The residence time on the site 1 is larger than the residence time at the site 2. On the other hand, for the backward transition the particle moves preferentially through the lane 2. That is why the forward transition is slower in this case. The situation is opposite for $w>1/4$ when there is a net counter-clockwise current. Only at equilibrium, $w=1/4$, the symmetry of the forward/ backward transition times is recovered \cite{berezhkovskii2019forward}.

In summary, we showed that the symmetry of forward/backward transition times could be broken even for a single particle. Using a random walk approach, it is  found that the transition times are the same only at equilibrium. In out-of-equilibrium, the direction of the loop current determines which transition times are shorter. This is the result of different contributions for forward and backward paths. Our analysis suggests that the asymmetry of forward/backward transition times is a measure of deviation from equilibrium \cite{battle2016broken}, and it reflects the complexity of underlying molecular mechanism.

%
%

\bibliography{main}
\bibliographystyle{plain}

\end{document}